\documentclass[times,twocolumn]{aastex7}
\hypersetup{urlcolor=magenta}

\usepackage{CJK}
\usepackage{enumitem}
\usepackage{amsmath}
\usepackage{multirow}
\usepackage{color}
\usepackage{bm}
 
\newcommand{\HL}[1]{#1}

\newcommand{\tE}{t_{\rm E}}

\newcommand{\e}{\varepsilon}

\DeclareGraphicsExtensions{.pdf,.png,.jpg}

\shorttitle{DREAMS DR1}
\shortauthors{Yang et al.}

\begin{document}
\begin{CJK*}{UTF8}{gbsn}
\renewcommand{\textsc}[1]{#1}
\title{{\large A Minute-Cadence Deep Bulge Survey: First Data Release of DREAMS}}

\author[0000-0003-0626-8465]{Hongjing Yang (杨弘靖)}
\affiliation{Department of Astronomy, Westlake University, Hangzhou 310030, Zhejiang Province, China}
\affiliation{Westlake Institute for Advanced Study, Hangzhou 310030, Zhejiang Province, China}
\email[show]{hongjing.yang@qq.com}
\email{yanghongjing@westlake.edu.cn}

\author[0000-0001-6000-3463]{Weicheng Zang (臧伟呈)}
\affiliation{Department of Astronomy, Westlake University, Hangzhou 310030, Zhejiang Province, China}
\email[show]{zangweicheng@westlake.edu.cn}

\author[0000-0001-5567-1301]{Francisco Valdes}
\affiliation{NSF NOIRLab, 950 N. Cherry Avenue, Tucson, AZ 85719, USA}
\email{frank.valdes@noirlab.edu}

\author[0000-0003-4625-8595]{Qiyue Qian}
\affiliation{Department of Astronomy, Westlake University, Hangzhou 310030, Zhejiang Province, China}
\affiliation{Department of Astronomy, Tsinghua University, Beijing 100084, China}
\email{qqy22@mails.tsinghua.edu.cn}

\author[0000-0001-5651-9440]{Yuchen Tang (唐雨辰)}
\affiliation{Department of Astronomy, Westlake University, Hangzhou 310030, Zhejiang Province, China}
\email{tangyuchen@westlake.edu.cn}

\author[0000-0002-1287-6064]{Zhixing Li (李知行)}
\affiliation{Department of Astronomy, Westlake University, Hangzhou 310030, Zhejiang Province, China}
\email{lizhixing@westlake.edu.cn}

\author[0009-0005-0410-8451]{Yuxin Shang (尚钰欣)}
\affiliation{Department of Astronomy, Tsinghua University, Beijing 100084, China}
\email{shangyx22@mails.tsinghua.edu.cn}

\author[0000-0001-8317-2788]{Shude Mao (毛淑德)}
\affiliation{Department of Astronomy, Westlake University, Hangzhou 310030, Zhejiang Province, China}
\email{shude.mao@westlake.edu.cn}

\author[0009-0003-7681-3702]{Yaosong Yu (于耀淞)}
\affiliation{Department of Astronomy, Westlake University, Hangzhou 310030, Zhejiang Province, China}
\email{ayuyaosong@gmail.com}

\author{Guillermo Damke}
\affiliation{Cerro Tololo Inter-American Observatory/NSF NOIRLab, Casilla 603, La Serena, Chile}
\email{guillermo.damke@noirlab.edu}

\author{Alfredo Zenteno}
\affiliation{Cerro Tololo Inter-American Observatory/NSF NOIRLab, Casilla 603, La Serena, Chile}
\email{alfredo.zenteno@noirlab.edu}

\author{Steve Heathcote}
\affiliation{Cerro Tololo Inter-American Observatory/NSF NOIRLab, Casilla 603, La Serena, Chile}
\email{steve.heathcote@noirlab.edu}

\author[0000-0003-4432-5037]{Konstantina Boutsia}
\affiliation{Cerro Tololo Inter-American Observatory/NSF NOIRLab, Casilla 603, La Serena, Chile}
\email{konstantina.boutsia@noirlab.edu}

\author[0009-0007-0032-4098]{Andong Xu (徐安东)}
\affiliation{Department of Astronomy, Westlake University, Hangzhou 310030, Zhejiang Province, China}
\email{xuandong@westlake.edu.cn}

\author[0009-0006-8010-4927]{Hao Ma (马皓)}
\affiliation{Department of Astronomy, Westlake University, Hangzhou 310030, Zhejiang Province, China}
\email{mahao@westlake.edu.cn}

\author[0000-0002-1279-0666]{Jiyuan Zhang (张纪元)}
\affiliation{Department of Astronomy, Tsinghua University, Beijing 100084, China}
\email{zhangjy22@mails.tsinghua.edu.cn}

\author[0009-0001-6584-7187]{Hongyu Li (李弘禹)}
\affiliation{Department of Astronomy, Tsinghua University, Beijing 100084, China}
\email{lihongyu25@mails.tsinghua.edu.cn}

\author[0000-0003-3201-061X]{Xikai Shan (单熙凯)}
\affiliation{Department of Astronomy, Tsinghua University, Beijing 100084, China}
\email{xk\_shan@mail.bnu.edu.cn}

\author[0000-0001-7016-1692]{Przemek Mr\'{o}z}
\affiliation{Astronomical Observatory, University of Warsaw, Al. Ujazdowskie 4, 00-478 Warszawa, Poland}
\email{pmroz@astrouw.edu.pl}

\author[0000-0002-7791-3671]{Xiurui Zhao (赵修瑞)}
\affiliation{Cahill Center for Astrophysics, California Institute of Technology, 1216 East California Boulevard, Pasadena, 91125, CA, USA}
\email{xiuruiz@caltech.edu}

\author{Andrew Gould} 
\affiliation{Max-Planck-Institute for Astronomy, K\"onigstuhl 17, 69117 Heidelberg, Germany}
\affiliation{Department of Astronomy, Ohio State University, 140 W. 18th Ave., Columbus, OH 43210, USA}
\email{gould.34@osu.edu}

\author[0000-0001-9481-7123]{Jennifer C. Yee}
\affiliation{Center for Astrophysics $|$ Harvard \& Smithsonian, 60 Garden St., Cambridge, MA 02138, USA}
\email{jyee@cfa.harvard.edu}

\author[0000-0003-0043-3925]{Chung-Uk Lee}
\affiliation{Korea Astronomy and Space Science Institute, Daejeon 34055, Republic of Korea}
\email{leecu@kasi.re.kr}

\author{Matthew Penny}
\affiliation{Department of Physics and Astronomy, Louisiana State University, Baton Rouge, LA 70803, USA}
\email{penny1@lsu.edu}

\author{Sean Terry}
\affiliation{Code 667, NASA Goddard Space Flight Center, Greenbelt, MD 20771, USA}
\affiliation{Department of Astronomy, University of Maryland, College Park, MD 20742, USA}
\email{skterry@umd.edu}

\author[0000-0003-2171-5083]{Patrick Tamburo}
\affiliation{Center for Astrophysics $|$ Harvard \& Smithsonian, 60 Garden St., Cambridge, MA 02138, USA}
\email{patrick.tamburo@cfa.harvard.edu}

\author{Tim Cunningham}
\affiliation{Center for Astrophysics $|$ Harvard \& Smithsonian, 60 Garden St., Cambridge, MA 02138, USA}
\email{timothy.cunningham@cfa.harvard.edu}

\author[0000-0002-4838-7676]{Quanzhi Ye (叶泉志)}
\affiliation{Department of Astronomy, University of Maryland, College Park, MD 20742, USA}
\affiliation{Center for Space Physics, Boston University, 725 Commonwealth Ave, Boston, MA 02215, USA}
\affiliation{Department of Earth, Planetary, \& Space Sciences, University of California, Los Angeles, CA 90095, USA}
\email{qye@umd.edu}

\author{Eric W. Peng}
\affiliation{NSF NOIRLab, 950 N. Cherry Avenue, Tucson, AZ 85719, USA}
\email{eric.peng@noirlab.edu}

\author{Rachel Street}
\affiliation{Las Cumbres Observatory Global Telescope Network, Inc., 6740 Cortona Drive, Suite 102, Goleta, CA 93117, USA}
\email{rstreet@lco.global}

\author{Katarzyna Kruszy{\'n}ska}
\affiliation{Las Cumbres Observatory Global Telescope Network, Inc., 6740 Cortona Drive, Suite 102, Goleta, CA 93117, USA}
\email{kkruszynska@lco.global}

\author{Etienne Bachelet}
\affiliation{Universit\'e Marie et Louis Pasteur, CNRS, Institut UTINAM UMR~6213, Besan\c{c}on, France}
\email{bachelet@ipac.caltech.edu}

\author{Yiannis Tsapras}
\affiliation{Astronomisches Rechen-Institut, M\"{o}nchhofstr. 12-14, D-69120 Heidelberg, Germany}
\email{ytsapras@ari.uni-heidelberg.de}

\author{Markus Hundertmark}
\affiliation{Astronomisches Rechen-Institut, M\"{o}nchhofstr. 12-14, D-69120 Heidelberg, Germany}
\email{markus.hundertmark@uni-heidelberg.de}

\collaboration{all}{(The DREAMS Collaboration)}

\correspondingauthor{Hongjing Yang, Weicheng Zang}

\begin{abstract}

The DECam Rogue Earths and Mars Survey (DREAMS), a NOIRLab survey program, has been conducting a three-year survey covering a 5\,deg$^2$ area in the Galactic bulge \HL{(roughly spanning $-1.2^\circ \lesssim \ell \lesssim +2.1^\circ$ and $-2.8^\circ \lesssim b \lesssim -0.6^\circ$)} since 2025 June. Its primary science goal is to detect low-mass free-floating planets through microlensing, while its minute-level cadence \HL{($20-40\,\mathrm{hr}^{-1}$ in $z$ band and $4-8\,\mathrm{hr}^{-1}$ in $r$ band)} also enables the detection and characterization of rapid phenomena on timescales of minutes to hours such as stellar flares and pulsating stars. \HL{The survey reaches a single-exposure depth of $z_{\rm AB}\sim 22$ mag, about two magnitudes deeper than previous bulge time-domain surveys.}
We present the data reduction and calibration of the DREAMS observations obtained in 2025 and introduce the first DREAMS data release (DR1). DR1 includes 1,856 $z$-band observations and 325 $r$-band observations for 59,372,789 stars. The DREAMS DR1 catalog contains \HL{about} twice as many stars as previous catalog covering the same 5 deg$^2$ area. We present DREAMS light curves for a known blue large-amplitude pulsator \HL{(BLAP)} and a known \HL{low-amplitude} transiting system to demonstrate the survey's capabilities. We also perform a pilot search for short-duration variables over about 0.4\% of the DR1 sample, identifying one new short microlensing event, two stellar flares, and 24 new short variables. This suggests that DREAMS DR1 may contain hundreds of stellar flares and thousands of previously unknown short variables.

\end{abstract}

\section{Introduction}\label{sec:intro}

Wide-field time-domain surveys have revolutionized our understanding of the dynamic universe by enabling systematic searches for transient and variable phenomena across large portions of the sky. The combination of wide angular coverage and repeated monitoring allows the detection of a large number of rare events which would be missed by targeted or small-field observations. Such surveys have uncovered a wide variety of transient and variable phenomena, including supernovae (e.g., \citealt{Nugent2011, Chen2024}), tidal disruption events (e.g., \citealt{Gezari2012, vanVelzen2020}), variable stars (e.g., \citealt{Soszynski2015,Udalski2018}), stellar flares (e.g., \citealt{Davenport2014,Gunther2020}), and eclipsing binaries (e.g., \citealt{Prvsa2011,Pawlak2016}), providing both population statistics and insights into the underlying physical processes.

Previously, most wide-field time-domain surveys have focused on monitoring large fractions of the sky to discover rare transients. These surveys typically cover areas of several thousand square degrees per night, enabling rapid identification of events across diverse stellar populations and extragalactic fields. For example, the Palomar Transient Factory (PTF, \citealt{PTF}) and its successor iPTF monitored thousands of square degrees per night with a typical cadence of 1--5 observations per night per field, while the Zwicky Transient Facility (ZTF, \citealt{ZTF}) scans the entire visible northern sky every few nights, achieving 2--6 observations per night for selected fields. The All-Sky Automated Survey for Supernovae (ASAS-SN, \citealt{ASASSN}) covers the full sky every 2--3 nights. Similarly, the Asteroid Terrestrial-impact Last Alert System (ATLAS, \citealt{ATLAS_Tonry2018}) scans the entire accessible sky every two nights.
Although these large-area surveys have been highly successful in discovering rare transients and building extensive time-domain catalogs, their relatively low cadence renders them insensitive to short-duration, non-periodic events occurring on timescales of minutes to hours.

Planet-hunting space-based missions, such as the \textit{Kepler} mission \citep{Borucki2010} and the Transiting Exoplanet Survey Satellite (TESS; \citealt{Ricker2015}), provide continuous, minute-level cadence photometry over durations ranging from about a month to several years. However, these observations are typically restricted to a limited set of pre-selected targets. For example, \textit{Kepler} monitored on the order of $10^5$ stars in its primary field, while TESS achieves 2-minute cadence mainly for selected targets, with the number of such objects also limited to $\sim10^5$. As a result, despite their excellent temporal resolution, these space-based surveys probe far fewer stars than wide-field ground-based programs, reducing their sensitivity to rare, non-periodic events occurring across large stellar populations.

Driven by the search for wide-orbit planets through microlensing \citep{MaoPaczyski1991,GouldLoeb1992}, high-cadence surveys have been conducted toward a $\sim 100\,{\rm deg}^2$ area of the Galactic bulge for the past three decades. The major programs are the second phase of the Microlensing Observations in Astrophysics survey (MOA-II, 2006--present; \citealt{Sako2008}), which employs a $2.2\,{\rm deg}^2$ camera on a 1.8 m telescope in New Zealand, the fourth phase of the Optical Gravitational Lensing Experiment (OGLE-IV, 2011--present; \citealt{OGLEIV}), which uses a $1.4\,{\rm deg}^2$ camera on a 1.3 m telescope in Chile, and the Korea Microlensing Telescope Network (KMTNet, 2016--present; \citealt{Kim2016_KMT}), which operates three 1.6 m telescopes, each equipped with a $4\,{\rm deg}^2$ camera, located in Chile, Australia, and South Africa. More recently, the PRime-focus Infrared Microlensing Experiment (PRIME, 2024--present; \citealt{PRIME_Kondo2023,PRIME_Sumi2025}) joined these efforts as the first dedicated near-infrared microlensing survey, utilizing a \(1.45\,{\rm deg}^2\) camera on a 1.8 m telescope in South Africa to monitor the high-extinction regions of the inner Galactic bulge. 
These surveys typically reach limiting magnitudes of $I\sim$19--20\footnote{For a signal-to-noise ratio of 5 with a single exposure.}.
Because typical microlensing signals last from several hours to about a day, these surveys operate at very high cadences, reaching up to $\Gamma \sim 6\,\mathrm{hr}^{-1}$ in their primary fields\footnote{Except for a $\sim 0.2\,{\rm deg}^2$ region of the KMTNet footprint that is covered by four overlapping prime fields, achieving a combined cadence of $\Gamma \sim 12\,\mathrm{hr}^{-1}$ from the KMTNet Australia and South Africa sites.}. These programs have discovered more than 280 wide-orbit planets \citep{NASAExo}, and recent statistical results suggest the presence of two distinct populations of planets (super-Earth/mini-Neptunes and gas giants) on wide orbits \citep{Zang2025Science_KMT_MassRatioFunction}.

Because of the dense stellar fields toward the Galactic bulge, these programs collectively monitor on the order of a few $10^8$ stars and are therefore also well suited for studies of short-duration variability. Previous investigations of variable phenomena in these fields have been based primarily on OGLE data led by the OGLE collaboration (e.g., \citealt{Soszynski2014}). However, light curves for the full stellar samples from these main programs are not publicly available. The ROME/REA survey, using the 1 m, $0.18\,{\rm deg}^2$ telescopes of the Las Cumbres Observatory Telescope Network (LCOGT, \citealt{LCOGT}), released light curves for $\sim 8 \times 10^6$ stars in the SDSS $g$, $r$, and $i$ passbands, but with a typical cadence of only $\Gamma \sim 1$--$2\,\mathrm{day}^{-1}$ \citep{Street2024}. These limitations affect systematic studies of short-duration variability toward the bulge. 

The Dark Energy Camera (DECam, \citealt{DECam2008,DECam2015}), mounted on the 4 m Blanco telescope at Cerro Tololo Inter-American Observatory (CTIO), provides a unique combination of wide field of view and large collecting area that is well suited for high-cadence, wide-field time-domain surveys. With a $3\,{\rm deg}^2$ field of view on a 4 m-class telescope, DECam enables monitoring of dense stellar fields to substantially greater depth than previous microlensing surveys conducted with 1-2 m facilities. In addition, its efficient operations allow rapid repeated imaging of the same fields, with overheads of 30 s between exposures and filter-change times of only $\sim 2$ s. These characteristics make DECam an ideal instrument for conducting deep, minute-cadence surveys over degree-scale fields while maintaining both high temporal resolution and large instantaneous sky coverage.

Beginning in June 2025, the DECam Rogue Earths And Mars Survey (DREAMS)\footnote{\url{https://time-allocation.noirlab.edu/\#/proposal/details/560332}} has been conducting a three-year survey covering a $5\,\mathrm{deg}^2$ area in the Galactic bulge. This program was initially motivated by the search for free-floating planets (FFPs) through microlensing, as the current constraints on the FFP population are limited by small-number statistics \citep{Mroz2017a,Gould2022_FFP_EinsteinDesert,Sumi2023_MOA9yr_FFP_MF}. DREAMS monitors $\sim 4\,{\rm deg}^2$ with cadences of $\Gamma_{z} = 20\,\mathrm{hr}^{-1}$ in the $z$ band and $\Gamma_{r} = 4\,\mathrm{hr}^{-1}$ in the $r$ band, while an additional $\sim 1\,{\rm deg}^2$ region is observed at twice these cadences. DREAMS reaches a limiting magnitude of $I\sim21.7$ ($z_{\rm AB}\sim22$), approximately 2 magnitudes deeper than previous bulge surveys. With its higher photometric precision and higher cadence, DREAMS is sensitive to Mars-mass FFPs in the Galactic bulge and Moon-mass FFPs in the Galactic disk \citep{KB251616}, and is expected to better characterize contamination from short-duration variables (e.g., stellar flares) in FFP signals \citep{mrozTESS,YangTESS}.

Beyond the search for FFPs, DREAMS also provides a powerful dataset for studies of short-duration variables. Its minute-level cadence enables the detection and characterization of rapid phenomena on timescales of minutes to hours, including, but not limited to, stellar flares, pulsating stars, and short-period eclipsing systems such as X-ray binaries. In addition, the larger aperture of the Blanco 4 m telescope allows DREAMS to reach significantly fainter magnitudes than previous bulge microlensing surveys. This increased depth opens a new discovery space for intrinsically faint and previously undetected variable sources in the crowded bulge fields, complementing earlier variability studies based on shallower, lower-cadence datasets.

To enable broader community use of this high-cadence, deep time-domain dataset, we have carried out the reduction and calibration of the DREAMS observations obtained in 2025 and present the first DREAMS data release (DR1), in collaboration with the Office of Information Technology and the High-performance Computing Center at Westlake University and the US NSF National Optical-Infrared Astronomy Research Laboratory (NSF NOIRLab). 
The DR1 includes 1,856 observations in the $z$ band and 325 observations in the $r$ band for 59,372,789 stars. Tools to query stars and extract individual star's information are provided. In this paper, we describe the observations (Section~\ref{sec:obs}), data processing and products (Sections~\ref{sec:data} and \ref{sec:release}) of DREAMS DR1. We also present DREAMS light curves for several short-duration variables and discuss the catalog depth in Section~\ref{sec:dis}.

\section{Survey Design and Observations}\label{sec:obs}
\begin{figure*}
    \centering
    \includegraphics[width=\linewidth]{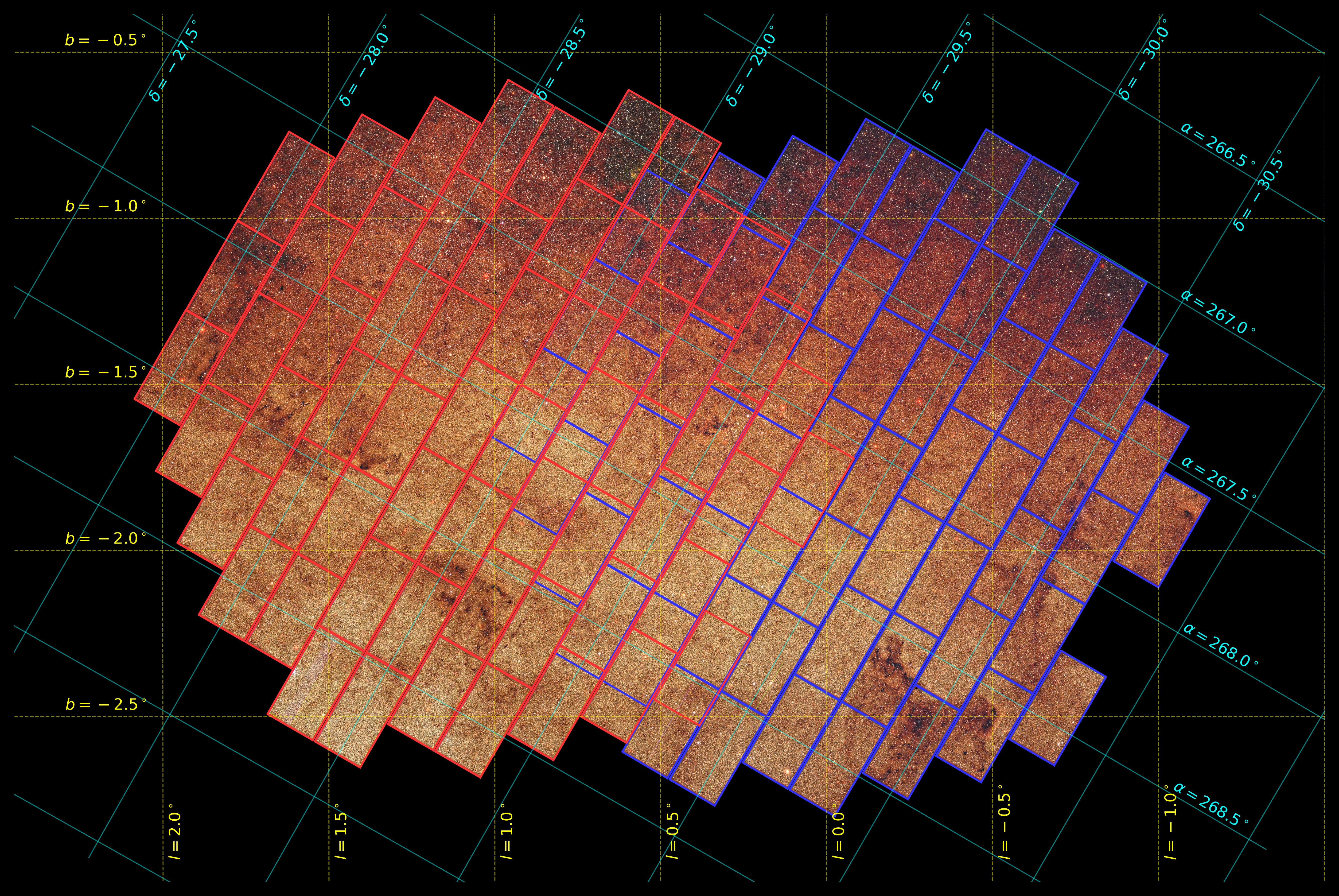}
    \caption{Sky map of the DREAMS field. 
    The image is in false RGB color composite from $z$-, $r$-, and $g$-band observations.
    The red and blue blocks mark the CCD layouts of the D01 and D02 fields, respectively.
    The Galactic coordinates $(l,b)$ and the Equatorial coordinates $(\alpha, \delta)$ are both labeled in yellow and cyan colors, respectively.
    An interactive version of the figure can be found on the DREAMS website (\url{https://astro.westlake.edu.cn/dreams/}).}
    \label{fig:skymap}
\end{figure*}

\subsection{Survey Design for DREAMS}

The primary driver of the DREAMS survey design is to maximize the detection efficiency for low-mass FFPs with masses $\lesssim 1\,M_{\oplus}$. According to the microlensing event-rate map derived from the OGLE-IV survey \citep{OGLEeventrate_bulge} and the event distribution reported by KMTNet \citep{Kim2018_KMTalgorithm}, the Galactic bulge region spanning $-1 \lesssim \ell \lesssim +2$ and $-3 \lesssim b \lesssim -1$ exhibits the highest event rate in the $I$ band. As shown in Figure~\ref{fig:skymap}, we place two DECam fields within this region, denoted D01 and D02, centerd at $(\alpha, \delta)_{\text{J2000}}$=(17:54:24, $-$29:00:00) and (17:53:00, $-$29:57:00), respectively. The total survey area is 5\,deg$^2$, with 1\,deg$^2$ covered by both fields, so this subregion is therefore observed at double cadence, optimized to enhance sensitivity to the lowest-mass FFPs accessible to DECam. The DREAMS field placement covers all low-extinction fields of the Roman Galactic Bulge Time Domain Survey \citep{MatthewWFIRSTI}, except for small regions that fall in the DECam CCD gaps, and overlaps with most of the microlensing survey area of the Earth~2.0 (ET) satellite \citep{Gould2021_FFP_ET,ET_WhitePaperv1}. No dithering is applied to these fields.

Because of the high extinction toward the Galactic bulge and the relatively high quantum efficiency of the DECam CCDs in the $z$ band\footnote{For DECam filter information, see \url{https://noirlab.edu/science/programs/ctio/filters/Dark-Energy-Camera}.}, $\sim 80\%$ of DREAMS images are taken in the $z$ band. To obtain color information for the variability, the remaining $\sim 20\%$ of the images are acquired in the $r$ band. 

The most common microlensing source stars in the OGLE, MOA, and KMTNet surveys are G dwarfs and subgiants. For such sources, the angular radius is typically $\theta_* \gtrsim 0.5\,\mu$as. Assuming a representative lens-source relative proper motion of $\mu_{\rm rel} = 6\,\mathrm{mas\,yr^{-1}}$, the source caustic-crossing time is $\gtrsim 2\theta_*/\mu_{\rm rel} = 1.5$\,hr. Because DREAMS probes deeper magnitudes, K dwarfs also become common source stars, for which the typical caustic-crossing time is closer to $\sim 1$\,hr. Adopting the FFP detection criteria used in the {\it Roman} FFP simulations \citep{Johnson2020}, namely, that at least six data points exceed the baseline flux by $\ge 3\sigma$, the required cadence for DREAMS is about $\Gamma = 6\,\mathrm{hr}^{-1}$. However, due to saturation limits, a single $z$-band exposure cannot exceed 60\,s. We therefore group several consecutive exposures (typically 3--5) into a single observing block for each field and plan to bin the data within each block when searching for FFPs. This strategy results in an effective minute-level cadence for DREAMS.

Because the Blanco telescope is operated in classical mode and observing time is typically allocated in half-night blocks, DREAMS observations are generally scheduled in half nights when the bulge is accessible for at least five hours, and in full nights when the bulge is accessible for at least nine hours, in order to maximize temporal coverage. As a result, the regular DREAMS observing season runs from April to August. Because the noise in the $z$ band is dominated by sky background from airglow and the blended light from field stars (rather than moonlight), most DREAMS observations are carried out during the bright and grey nights.

The DREAMS team also accesses observing time on the 3.6\,m Canada-France-Hawaii Telescope (CFHT) and the 8.2\,m Subaru Telescope through their partner institutes to obtain simultaneous observations with DECam, as well as consecutive coverage following the Chilean night. The Subaru observations are conducted in the $z$ band, while the CFHT images are taken in the $r$ band to complement the color information.

\subsection{Observations in 2025}

\begin{figure*}
    \centering
    \includegraphics[width=0.8\linewidth]{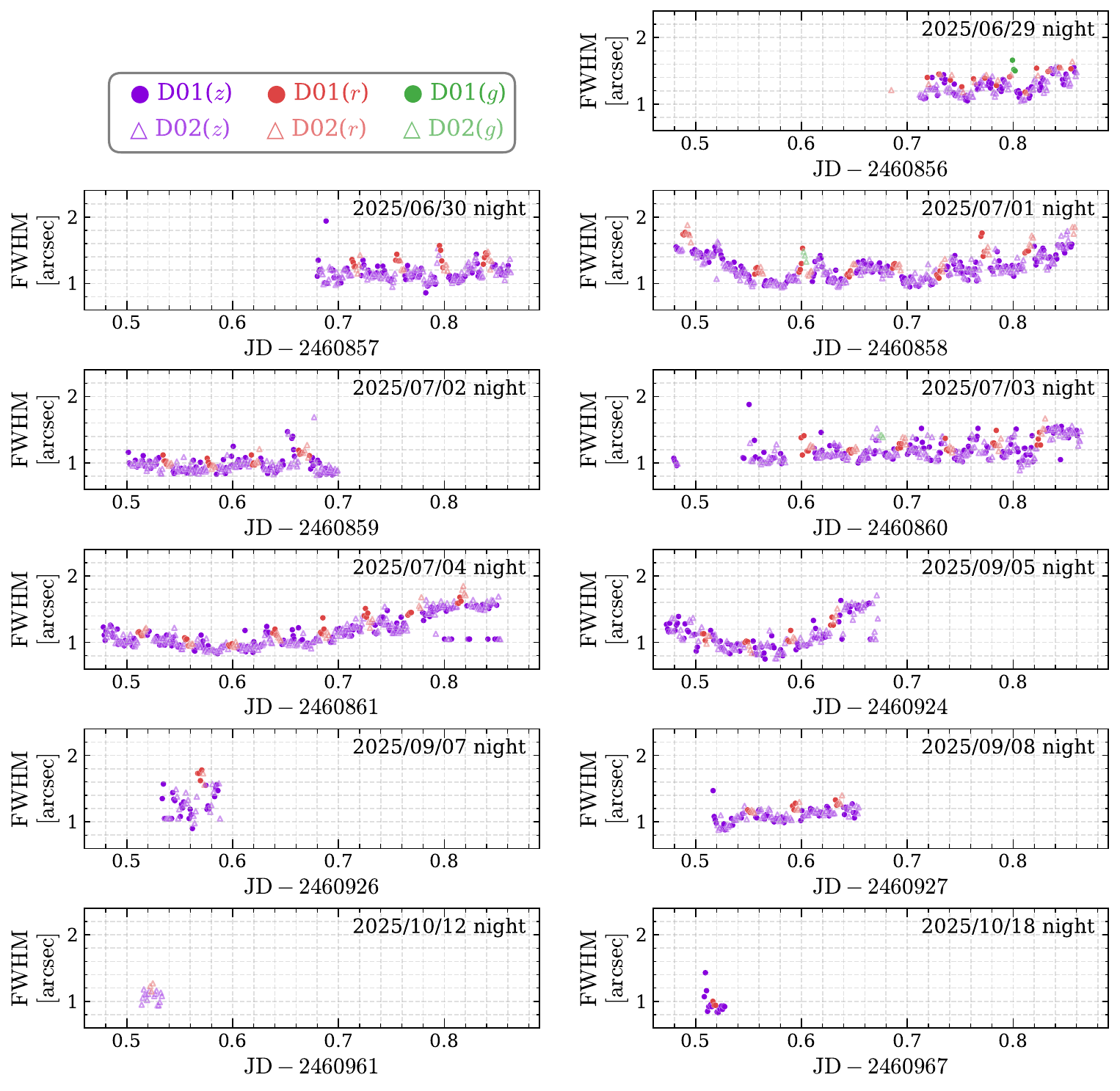}
    \caption{Full-width half-maximum (FWHM) of the point spread function (PSF) of the images as a function of time for each observation night in 2025. The $z$-, $r$-, and $g$-band observations are colored in magenta, red, and green, respectively. D01 and D02 observations are marked in circles and triangles, respectively.}
    \label{fig:obs2025}
\end{figure*}

The DECam observations in the 2025A season (February to July) were conducted under the NOIRLab proposal 2025A-80629. The first allocated night was June 15, but the entire night was lost due to clouds. Subsequent observations were carried out over six nights from June 29 to July 4. Among these, July 1, 3, and 4 each included an approximately 9\,hr block of DREAMS observations, although a small fraction of the time was affected by clouds and technical issues. 

In late July and early August, a heavy snowstorm struck CTIO. As a result, the final night of the 2025A season and the first two nights of the 2025B season, under the new NOIRLab survey proposal 2025B-56033 for DREAMS, were lost. Additional DECam observations in August were canceled due to a fracture in the rail on which the Blanco dome rides. Through Director's Discretionary Time (DDT), observations were obtained on September 5, 7, and 8, with nightly coverage ranging from 1.3\,hr to 4.9\,hr. On October 12 and 18, additional 0.5\,hr observations were obtained for the D02 and D01 fields, respectively, to provide baseline flux measurements for long microlensing events and to improve the source color measurement.

CFHT obtained $r$-band images for most DREAMS fields using the 1\,deg$^2$ imager MegaCam \citep{MegaCam} on June 29, 30, and July 1. On each night, the CFHT observations overlapped with the DECam observations for about 100\,min, followed by an additional $\sim 1$\,hr. The CFHT data have not yet been reduced and will be released as part of DREAMS DR2. No Subaru observations were carried out in 2025.

For the 2025A season, the DECam strategy was to sample the D01 and D02 fields once per hour using five $z$-band blocks and one $r$-band block. Each block comprised four exposures, with individual exposure times of 42\,s in $z$ and 60\,s in $r$. Based on a visual inspection at the time, the level of saturation did not appear to be severe. Therefore, for the 2025B season, the individual exposure times were increased to 60\,s in $z$ and 80\,s in $r$, with each hour consisting of four $z$-band blocks and one $r$-band block. Each $z$-band block contained four exposures, while each $r$-band block contained three exposures. In addition, three and six $g$-band exposures were obtained for the D01 and D02 fields, respectively, to further constrain the color of the baseline object.

However, the end-of-year data reduction revealed two issues: (1) the 60\,s $z$-band exposures suffered from significant saturation, and (2) the effective cadence of $r$-band observations was only about one hour, limiting color information for short-duration events. To address these, we revised the strategy for future seasons (from 2026 onward) as follows: the $z$-band exposure time is reduced back to 42\,s to avoid saturation, and the $r$-band exposure time remains 80\,s. The new cadence is increased by interleaving $r$ exposures between $z$ blocks, achieving a cycle time of $\sim$15\,min. Specifically, for each field the sequence is: three $z$ exposures (42\,s), one $r$ exposure (80\,s), followed by two $z$ exposures (42\,s). This sequence is executed alternately for D01 and D02 throughout the night. Table~\ref{tab:obs_strategy} summarizes the evolution of the observing strategy.


\begin{table*}
    \centering
    \caption{Summary of DECam observing strategies for DREAMS}
    \label{tab:obs_strategy}
    \begin{tabular}{lcccc}
    \hline\hline
    \multirow{2}{*}{Season} & $z$-band & $r$-band & \multirow{2}{*}{Observational cycle} & Time cost \\
                            & exposure & exposure &                                      & per cycle \\
    \hline
    2025A  & 42\,s & 60\,s & $(4~z_{\rm D01},4~z_{\rm D02})\times5$, $(4~r_{\rm D02},4~r_{\rm D01})\times1$ & 59.7\,min \\
    2025B  & 60\,s & 80\,s & $(4~z_{\rm D01},4~z_{\rm D02})\times4$, $(3~r_{\rm D02},3~r_{\rm D01})\times1$ & 58.0\,min \\
    2026$+$& 42\,s & 80\,s & $(3~z,1~r,2~z)_{\rm D01}$, $(3~z,1~r,2~z)_{\rm D02}$ & 15.5\,min \\
    \hline
    \end{tabular}
\end{table*}

After visually reviewing and removing images with issues, the DREAMS DR1 contains a total of (929, 927) $z$-band data points and (164, 161) $r$-band data points for the (D01, D02) fields. Figure~\ref{fig:obs2025} summarizes the distributions of the full-width half-maximum (FWHM) of the point spread function (PSF) for DR1 across the 11 observing nights. The best PSF FWHM values are 0.72$''$ in $z$ and 0.80$''$ in $r$.

\section{Data Processing}\label{sec:data}

\subsection{Preamble}

To extract high-precision photometry from the crowded DREAMS fields, we employ difference image analysis (DIA) \citep{Tomaney1996, Alard1998, Alard2000, DanDIA}. This technique is well established for time-domain surveys in dense fields and has proven powerful in many Galactic bulge microlensing surveys (e.g., \citealt{Albrow1998PLANET, Wozniak2000_diapl, Bond2001, Yang2024_pysis5_RAMP1}).

We have developed a modified version of the KMTNet pySIS pipeline \citep{pysis, Yang2024_pysis5_RAMP1, Yang2025_RAMP2, Qian2025_KMTFFP1} specifically tailored to the DECam full-frame images. 
Our modifications primarily aim to fully exploit the products delivered by the DECam community pipeline \citep{DECampipeline}, including the calibrated images, weight maps, and mask maps, and to handle the specific characteristics of DECam images, such as the large number of CCDs. 

The overall data processing workflow consists of the following  steps, each described in detail in the subsequent subsections. 
First, we preprocess the individual DECam exposures to measure image quality metrics and then cut the full-frame images into small stamps (Section~\ref{sec:pipeline:preprocess}). 
Second, we construct a high-quality reference image for each field and band, which serves as the template for image subtraction (Section~\ref{sec:pipeline:reference}).  
Third, we perform image subtraction between each science image and the reference image using a spatially varying, pixelated convolution kernel \citep{DanDIA,pysis}, producing difference images (Section~\ref{sec:pipeline:imsub}). 
Fourth, we construct an input star catalog from the master reference image, containing the positions for all detectable sources (Section~\ref{sec:pipeline:catalog}).
Finally, we extract PSF photometry at the positions of all catalog stars from each difference image, yielding time-series light curves (Section~\ref{sec:pipeline:lc}).

\subsection{Preprocessing} \label{sec:pipeline:preprocess}
The raw DECam images are processed by the DECam community pipeline \citep{DECampipeline}, which performs standard calibrations (bias, dark, flat-fielding) and astrometric solution. For each exposure, the pipeline provides a calibrated science image, a weight map (inverse variance), a mask map (flagging bad pixels, cosmic rays, saturation, etc.), and World Coordinate System (WCS) information with typical astrometric residuals of $\lesssim1$\,pix (1\,pix $\approx 0.263''$).
These products serve as the starting point for our DREAMS data reduction.

Before proceeding, we visually inspect all images to remove those affected by tracking errors, defocusing, or other obvious artifacts, ensuring that only good-quality data enter the subsequent processing. As a result, 11 and 3 images are removed for D01 and D02 field, respectively. Among them, 2 are caused by defocusing and 12 are due to tracking errors.

Because the PSF varies significantly across the entire field of view, the images are then divided into small stamps before further processing.
We divide each DECAM CCD into a fixed grid of small cutouts. Because the DREAMS observations cover two fields (D01 and D02) with multiple epochs, we first establish a common astrometric reference frame for each band and field. For a given band and field, we compute the median pointing offset from all available exposures and select the image closest to this median pointing as the astrometric reference image. This reference image defines the sky-to-pixel mapping used to cut all other images consistently.

DECam has 61 available CCD chips (detectors). For each CCD chip ($2048 \times 4096$ pixels), we lay out a grid of $4 \times 8$ stamps, yielding 32 stamps per chip. Each stamp is a $600 \times 600$ pixel square, centered at a predefined sky position. The grid ensures neighboring stamps overlap by 82 pixels, providing continuous coverage with sufficient redundancy. This overlap ensures that sources near the edge of one stamp fall closer to the center of a neighboring stamp, allowing reliable photometry. Stamp centers are expressed in equatorial coordinates ($\alpha, \delta$), which are derived from the astrometric reference image's WCS.

For every science exposure (with its corresponding weight and mask maps), we process each CCD chip independently. Using the WCS from the DECam pipeline, we convert each stamp's predefined sky coordinates to pixel coordinates and extract a $600 \times 600$ pixel cutout around that center. If the cutout extends beyond the CCD boundaries, we pad the missing region with zeros for the science image and weight map, and with ones (indicating ``bad'' or ``no data'') for the mask map. This padding guarantees that every stamp from every exposure has the same dimensions and is aligned to the same sky region, despite sub-pixel level dithers or pointing variations, which can be well handled by our DIA algorithms. The cutout image, weight map, and mask map are saved together, with updated header keywords recording the original center position, cutout boundaries, and preserving the WCS information.

For each exposure, we also measure representative image quality parameters using a fixed stamp near the field center (e.g., stamp 0402 on chip S4). These include the PSF FWHM, PSF ellipticity (a measure of irregularity), sky background, and the standard deviation of the entire image. These quality metrics are used to monitor observing conditions and can inform the later reference image selection.

All cutouts are organized by field, stamp identifier, and band (e.g., \texttt{D01.S1.0101\_z}, where \texttt{D01} is the field, \texttt{S1} the CCD chip, \texttt{0101} the stamp ID, and \texttt{z} the band). The resulting stamp images are then ready for subsequent processing steps.

\subsection{Reference Image Construction} \label{sec:pipeline:reference}

\begin{figure*}
    \centering
    \includegraphics[width=1.0\linewidth]{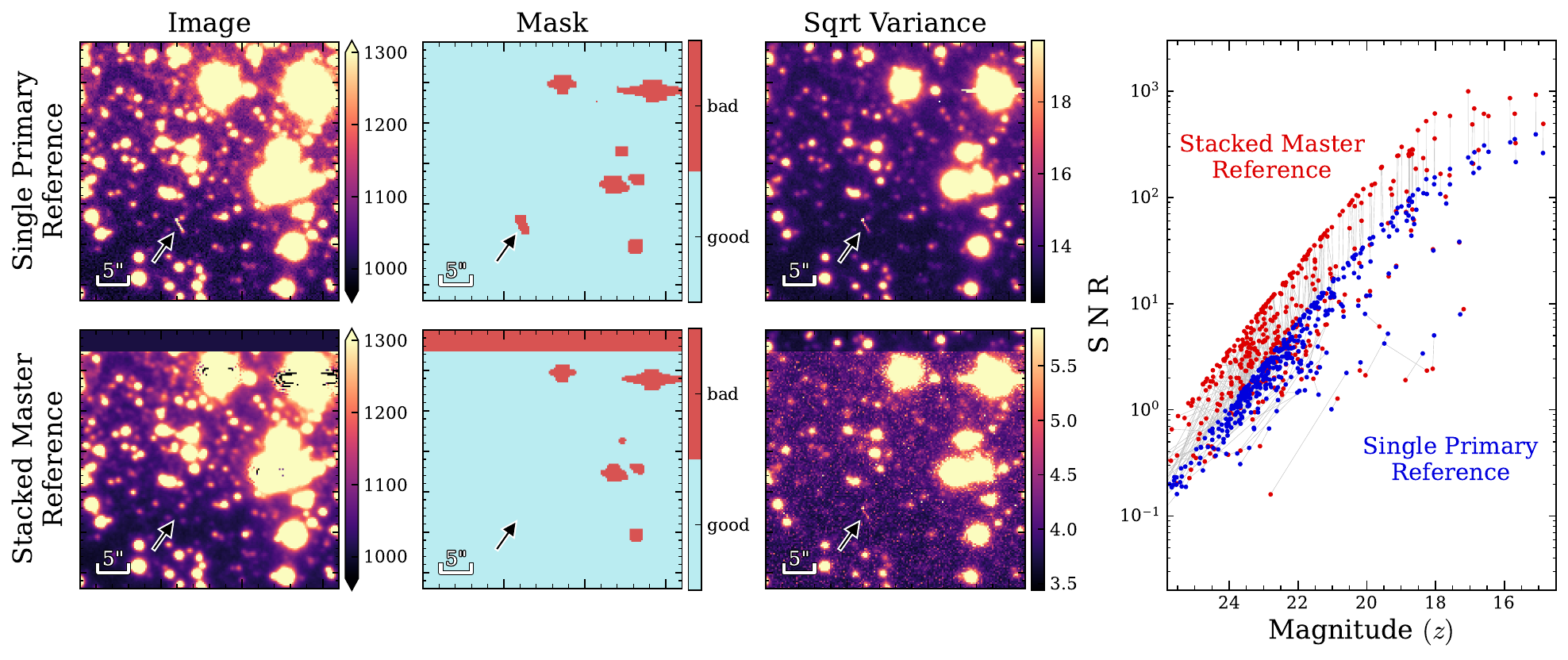}
    \caption{Comparison between the single-exposure primary reference (top row) and the stacked master reference (bottom row) images \HL{in $z$ band}. \HL{The left three columns show} the image, mask, and noise map for a $0.7'\times0.7'$ example region in the D01 field, centered at $(\alpha, \delta)\sim$(17:50:25, $-$28:47:23). The arrow indicates a cosmic ray present in the primary image but removed after stacking. All \HL{image} panels are oriented with north up and east left. \HL{The pixel scale is $0.263''$.} \HL{The rightmost column shows the measured signal-to-noise ratio (SNR) and magnitude for the same set of sources from the single exposure (blue) and in the stacked reference image (red), respectively. Identical sources are connected to illustrate the precision improvement from stacking.}}
    \label{fig:ref}
\end{figure*}

A high-quality reference image is essential for difference image analysis, as it serves as the template against which all science images are subtracted. For each band and field, we construct a stacked master reference image from a set of the best-seeing exposures.

We first select candidate reference images based on three quality metrics measured during preprocessing (Section~\ref{sec:pipeline:preprocess}): the PSF FWHM (see Figure~\ref{fig:obs2025}), the PSF ellipticity, and the sky background. 
Images with small FWHM, low ellipticity (i.e., round PSF shapes), and low background are preferred. From these candidates, we visually inspect and select 20 images with consistently good quality for each band and field. Among this set, we designate one image as the ``primary reference''. The remaining 19 images are then convolved, using the same algorithm detailed in Section~\ref{sec:pipeline:imsub}, to match the PSF of the primary reference image. This ensures that all images contributing to the stack share a common PSF and instrumental flux scale. 
To minimize the flux scale discrepancy between the two fields, for each band, we select two adjacent exposures, one from D01 and one from D02, as the ``primary references'' for their respective fields.

Throughout the convolution and stacking process, we consistently propagate the corresponding weight and mask maps. The stacking process employs a sigma-clipping weighted average on a pixel-by-pixel basis. For each pixel, we classify it as follows: (a) if five or more valid images are available, we perform iterative sigma-clipping to reject outliers and compute the weighted mean; (b) if fewer than five valid images exist, we compute the weighted mean without sigma-clipping and mark the pixel as ``bad'' in the mask; (c) pixels with no valid data are set to zero in the science image and flagged as ``bad'' in the mask.

This procedure yields a deep, high signal-to-noise master reference image for each band and field, along with co-added weight and mask maps, as shown in Figure~\ref{fig:ref}. 
The master reference image is then used for star catalog construction (Section~\ref{sec:pipeline:catalog}) and as the template for all subsequent image subtractions (Section~\ref{sec:pipeline:imsub}).

\subsection{Image Subtraction} \label{sec:pipeline:imsub}

\begin{figure*}
    \centering
    \includegraphics[width=0.8\linewidth]{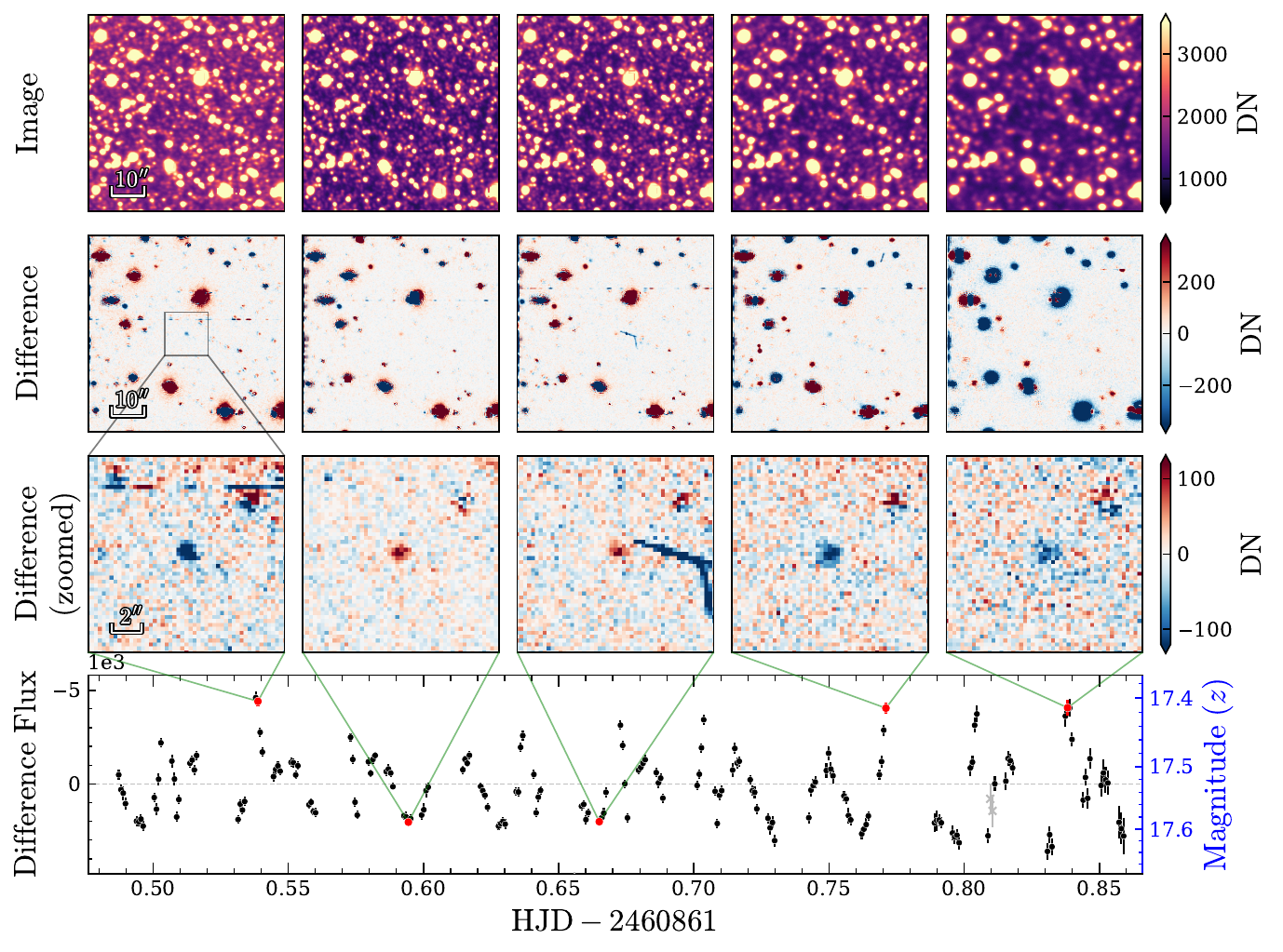}
    \caption{Demonstration of image subtraction near a blue large-amplitude pulsator (BLAP), OGLE-BLAP-019 \citep{Pietrukowicz2025_BLAP_OGLE}, observed in the $z$ band in the D02 field, with DREAMS identifier 45011997. The five columns correspond to selected epochs from the night of 2025 July 5. {\it Top row:} Science images. {\it Middle row:} Difference images. {\it Third row:} Zoomed-in views of the difference images centered on OGLE-BLAP-019. The color scale is fixed within each row. {\it Bottom panel:} Difference flux light curve; gray points are flagged as bad measurements, and the five highlighted red points mark the selected epochs. All fluxes are in digital number \HL{(DN)} units.}
    \label{fig:diff}
\end{figure*}

Image subtraction is performed using the pySIS pipeline \citep{pysis, Yang2024_pysis5_RAMP1, Yang2025_RAMP2}, which implements a spatially varying, pixelated convolution kernel (for more details about the algorithm, see \citealt{DanDIA} and \citealt{pysis}). For each stamp, we solve the convolution kernel, convolve the target image to match the PSF and flux scale of the master reference image, and then subtract to obtain the difference image. 

Throughout the process, we propagate the input weight (variance) and mask maps from both the reference and target images, ensuring that pixel uncertainties and bad pixels are properly accounted for in the final difference image and its associated variance map.

We make several modifications to adapt the pipeline to DREAMS data. These include: (1) incorporating both reference and target weight and mask maps into the kernel-solving and subtraction steps; (2) fine-tuning some key parameters, such as kernel size, spatial variation order, number of reference stars, and outlier rejection thresholds, to achieve optimal performance in the DECam images.

The resulting difference images, variance maps, and masks provide the basis for photometric extraction described in Section~\ref{sec:pipeline:lc}. Additionally, we compute a subtraction quality metric $\sigma_\mathrm{subt}$ \citep{Yang2024_pysis5_RAMP1} for each subtraction, which will be used to flag unreliable measurements in the final light curves.

Figure~\ref{fig:diff} shows an example of the image subtraction process. The difference images are overall clean. They show residuals primarily from saturated stars, along with additional residuals from bad pixels (visible in the first column), cosmic rays (visible in the third column), and from the variable stars (in this example, the BLAP target itself). Applying PSF photometry (Section~\ref{sec:pipeline:lc}) to these difference images could yield (difference) flux measurements.

\subsection{Input Catalog} \label{sec:pipeline:catalog}

A complete and accurate input star catalog is required for forced photometry on the difference images. In the DREAMS fields, our DECam images reach significantly deeper than existing public catalogs (e.g., OGLE \citep{OGLEIII}, Gaia \citep{GaiaDR3_2023_summary}, and DECaPS2 \citep{Saydjari2023ApJS_DECaPS2}), rendering external catalogs insufficient for our purposes. We therefore construct the catalog directly from the DECam images, using the deep, high signal-to-noise $z$-band master reference images described in Section~\ref{sec:pipeline:reference}.

\subsubsection{Catalog Extraction}
Source catalogs are extracted independently for each stamp using \textsc{DoPHOT} \citep{dophot}. For each stamp, we provide the master reference image and the corresponding mask map (\textsc{DoPHOT} does not support inputting a weight map). \textsc{DoPHOT} iteratively detects pixels above a threshold, fits them with an analytic PSF model, and subtracts the fitted sources from the image. The detection threshold starts high and decreases to a final signal-to-noise ratio (SNR) of 5 for the single central pixel, where the noise level is estimated internally by \textsc{DoPHOT} from the combination of the sky background Poisson noise and the readout noise. 
For each detected source, \textsc{DoPHOT} performs an elliptical Gaussian-like PSF fit to measure its position, flux, and flux error. Detected sources are further classified based on their shape into stars, galaxies, cosmic rays, artifacts, etc. Because we expect only stars in our fields and all cosmic rays are already eliminated during reference image construction (Section~\ref{sec:pipeline:reference}), we only exclude artifacts reported by \textsc{DoPHOT}, which are dominated by false objects around saturated pixels.

After running \textsc{DoPHOT} over all stamps, a total of 76,197,680 sources are initially extracted. However, because each stamp is processed independently and neighboring stamps overlap, the resulting catalogs contain duplicates within each chip.
Moreover, the two DREAMS fields (D01 and D02) share an overlapping region of approximately 1\,deg$^2$, which further introduces duplicates between fields.
Consequently, a multi-level merging procedure is required to obtain a clean, non-redundant input catalog for the subsequent photometry.

\subsubsection{Catalog Merging}

We merge the stamp-level catalogs in three hierarchical steps: chip-level, field-level, and global.

\textbf{Chip-level merging.} For each CCD chip, we divide the detector plane into 32 non-overlapping regions, each associated with one stamp, by assigning every position to the nearest stamp center. For each stamp, we remove sources whose positions fall outside its own region, i.e., those that belong to the region of a neighboring stamp. The remaining sources from all stamps are then combined, producing a complete and duplicate‑free catalog for each chip.

\textbf{Field-level merging.} Within a given field (D01 or D02), the 61 chips are non-overlapping on the sky. The chip-level catalogs are therefore simply concatenated to form a field-wide source catalog. No further de-duplication is required at this stage. The D01 field contains 33,619,488 sources and D02 field 30,964,846 sources.

\textbf{Cross-field matching.} The two fields, D01 and D02, overlap by about 1\,deg$^2$. To obtain a single global catalog, we cross-match the field catalogs using a mutual-nearest-neighbor algorithm, requiring a match to be within $0.263''$ (1 pixel) and instrumental magnitude difference $|\Delta m| \le 0.5$ mag. Matched sources are merged, retaining information from both fields; unmatched sources from either field are kept with the other field's information set to null.

The final global catalog contains positions and instrumental magnitudes for 59,372,789 sources and serves as the positional reference for forced photometry in Section~\ref{sec:pipeline:lc}. 
The cross‑matching reduced the total source number from $\sim64.6$ million by $\sim5.2$ million, suggesting that about 40\% of the expected duplicates in the overlapping region were successfully merged.
However, a substantial number of duplicates likely remain in the overlapping area, primarily among fainter stars where positional and photometric uncertainties are larger.
Mismatches (incorrectly merging distinct stars) are expected to be rare.
These issues are expected to be resolved in future releases by incorporating complete, deeper, and higher-resolution reference observations (Section~\ref{sec:dis:catalog}).

\subsection{Light Curve} \label{sec:pipeline:lc}

We extract forced PSF photometry for every source in the input catalog (Section~\ref{sec:pipeline:catalog}) from both the difference images and the reference images, following the method detailed in \citet{Yang2024_pysis5_RAMP1} and properly incorporating the weight and mask maps.

For each stamp and band, a high SNR PSF model is constructed from the master reference image by median stacking several tens to hundreds of bright, isolated stars. 
This reference PSF is then convolved with the spatially varying kernel derived from the image subtraction step (Section~\ref{sec:pipeline:imsub}) to obtain a PSF model that matches the target image. 
At each source position, we perform a linear fit (with iterative sigma clipping) of this convolved PSF model to the local pixels of the difference image, yielding the difference flux $F_\mathrm{diff}$ (reference minus target), its uncertainty, and its quality metric $\sigma_\mathrm{phot}$ \citep{Yang2025_RAMP2}. 
The same fit applied to the reference image itself, using the unconvolved reference PSF, gives the baseline flux $F_\mathrm{base}$, its error, and its own $\sigma_\mathrm{phot}$. 

Instrumental fluxes are converted to calibrated magnitudes using zero points determined from cross-matching with the DECaPS2 catalog \citep{Saydjari2023ApJS_DECaPS2}, which provides deep photometry in the $g, r, i, z, y$ bands on the AB magnitude system. 
For each chip and band, we identify bright stars common to both catalogs ($z_\mathrm{AB}\lesssim19$) and compute an iterative sigma-clipped weighted mean offset between the DECaPS2 magnitudes and our baseline instrumental magnitudes.
Because the primary science goal, microlensing event search, is insensitive to the absolute zero point, the resulting zero points are only approximately calibrated (typical uncertainty $\lesssim0.05$\,mag) and serve to roughly convert the light to magnitudes. More accurate calibration can be achieved in the future using, for example, red clump stars, or by interested investigators for applications requiring higher precision \citep[e.g.,][]{Nataf2013}.

The final light curves from all 2025 observations are stored per stamp and per band in HDF5 format. 
Each HDF5 file is $\sim520$\,MB for the $z$ band and $\sim150$\,MB for the $r$ band, with a total volume of $\sim2.5$\,TB.
Each HDF5 file is self-describing and contains the following information: the source catalog with positions and identifiers, WCS information, baseline flux measurements (which, if combined for both bands, allow construction of color-magnitude diagrams), per-exposure difference fluxes with uncertainties and photometric quality metrics, the reference image itself (useful as a finding chart), and image quality metrics. 
This dataset provides the foundation for all future variability studies using DREAMS.

\subsection{Photometric Accuracy Performance}

\begin{figure}
    \centering
    \includegraphics[width=1.0\linewidth]{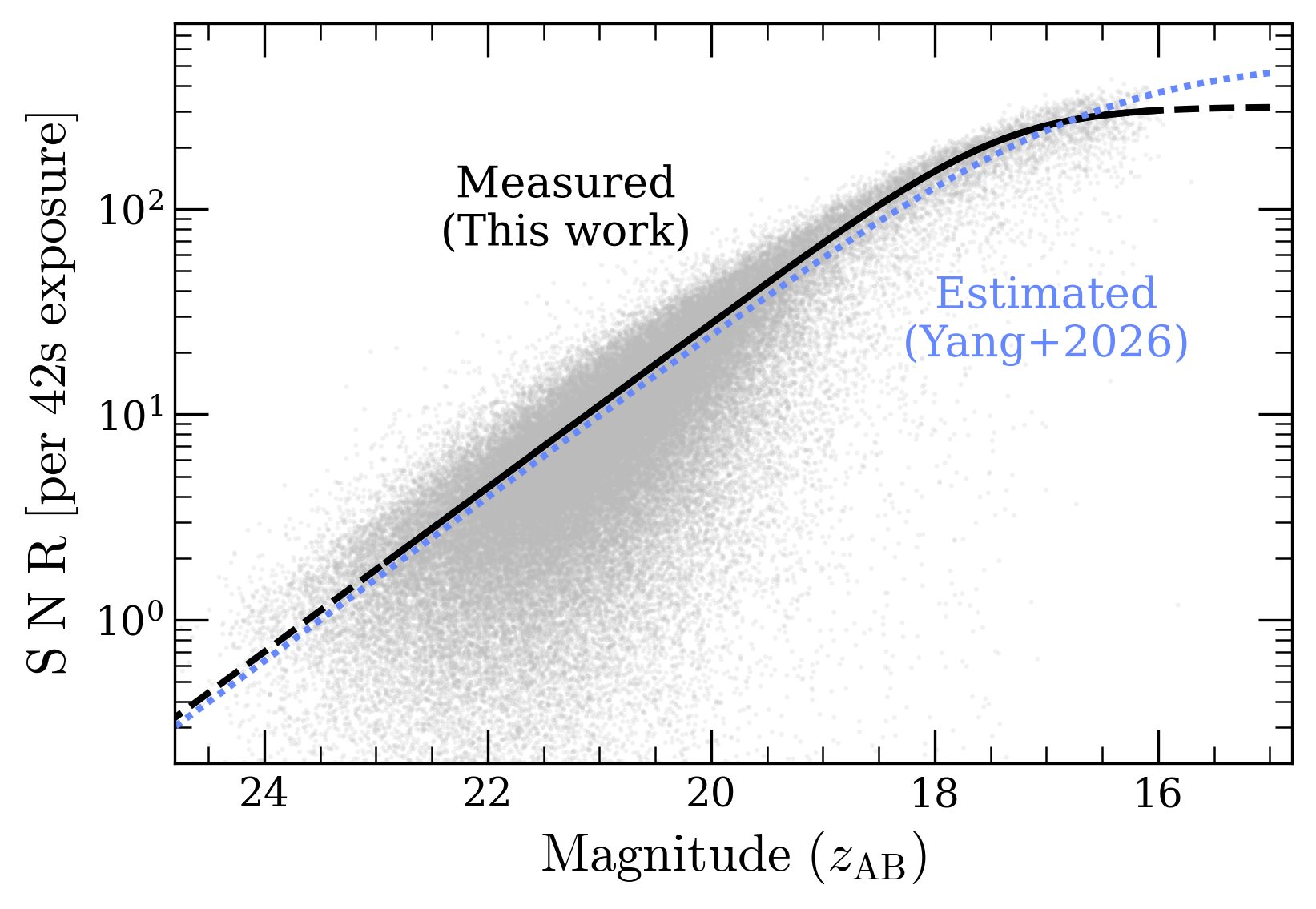}
    \caption{Signal-to-noise ratio as a function of $z$-band magnitude for individual 42\,s exposures. Gray points show a random subset of $10^6$ stars. The black line is the median SNR curve. For comparison, the blue dotted line shows the SNR estimate by \citet{KB251616}.}
    \label{fig:snr}
\end{figure}

To characterize the photometric performance of the DREAMS DR1 data, we derive the per-exposure SNR as a function of magnitude for the $z$-band observations (42\,s exposure time). For each source, we compute the SNR as the baseline flux divided by the root-mean-square (RMS) scatter of its light curve. The RMS includes both noise and intrinsic source variability, but for the majority of non-variable stars, it serves as a reliable measure of the typical photometric uncertainty. We bin all sources by magnitude (0.1\,mag bin width) and compute the median SNR per bin after iterative $1.5\sigma$ clipping to remove outliers, including primarily variable stars whose RMS is inflated by real variability.

Figure~\ref{fig:snr} shows the resulting SNR curve, with a random subset of $10^6$ sources plotted in the background to illustrate the distribution. 
As expected, the SNR increases for brighter sources. However, it plateaus around $z_{\rm AB}\sim17$, well before the hardware saturation limit ($z_{\rm AB}\sim15$ for 42\,s exposures). This is due to systematic errors, e.g., non-linearity, PSF model inaccuracy, flat-field residuals, that dominate the noise rather than the photon noise.
\HL{Nevertheless, the photometry remains reliable for $17\gtrsim z\gtrsim15$ and the plateau simply implies that the precision no longer improves for brighter sources in this regime.}
For comparison, we also overplot the estimated DREAMS SNR given by \citet{KB251616}, which was used to predict DREAMS's sensitivity to low-mass free-floating planets. The measured SNR of the actual data slightly exceeds the estimation, confirming that DREAMS reaches the expected sensitivity.

\section{Released Data Products}\label{sec:release}
We present the first data release (DR1) of the DREAMS project, comprising high-cadence light curves for approximately 59 million sources toward the Galactic bulge, observed in the $z$ and $r$ bands during the 2025 season. All data products are publicly accessible through the DREAMS data portal:
\begin{center}
    \textbf{\url{https://astro.westlake.edu.cn/dreams/}}
\end{center}

\subsection{Data Products Overview}
For each of the $\sim$59 million sources, DR1 provides the following data products:
\begin{itemize}
    \item \textbf{Source catalog entry}: coordinates, approximately calibrated instrumental magnitudes in $z$ and $r$ from the baseline (stacked master reference image).
    \item \textbf{Finding chart}: a $1'\times1'$ cutout from the deep $z$-band master reference image, centered on the source.
    \item \textbf{Color-magnitude diagram (CMD) information}: approximately calibrated instrumental color and magnitudes of all stars in the same stamp ($\sim 2.6'\times2.6'$), enabling placement of the source in a $z$ versus $r-z$ CMD. 
    \item \textbf{Light curves}: time-series photometry for all available epochs in the 2025 season, provided in both instrumental flux and approximately calibrated instrumental magnitude (Section~\ref{sec:pipeline:lc}). Light curves include per-epoch fluxes, uncertainties, image quality metrics, and photometric quality flags.
\end{itemize}

\begin{table*}[htbp]
    \centering
    \caption{Data product summary for a single DREAMS source}
    \begin{tabular}{l|l|l|l}
        \hline\hline
        \multicolumn{4}{c}{\textbf{Light curve}} \\
        \multicolumn{4}{c}{\textit{(Each row corresponds to a single exposure, each field and band is in an individual file)}} \\
        \hline
        Parameter & Description & Parameter & Description \\
        \hline
        image\_id   & Internal image identifier            & stddev      & Image digital number standard deviation \\
        mjd         & Modified Julian Date at mid-exposure & airmass     & Airmass of the observation \\
        flux        & Instrumental flux (digital number)   & sigma\_subt & Image subtraction quality (smaller better) \\
        ferr        & Instrumental flux uncertainty        & sigma\_phot & PSF photometry quality (smaller better) \\
        fwhm        & PSF FWHM (pixels)                    & mag         & Approximately calibrated magnitude (AB) \\
        sky         & Sky background                       & merr        & Magnitude uncertainty \\
        qirr        & PSF irregularity \citep{Yang2024_pysis5_RAMP1}  & flag        & Quality flag (0 = good) \\
        \hline\hline
    
        \multicolumn{4}{c}{\textbf{CMD data}} \\
        \multicolumn{4}{c}{\textit{(Each row corresponds to a single star in the same stamp as the target.)}} \\
        \hline
        Parameter & Description & Parameter & Description \\
        \hline
        r\_x    & X coordinate (pixel) on the $r$-band image      & z\_x    & X coordinate (pixel) on $z$-band image\\
        r\_y    & Y coordinate (pixel) on the $r$-band image      & z\_y    & Y coordinate (pixel) on $z$-band image \\
        r\_ra   & Right Ascension measured on the   & z\_ra   & Right Ascension measured on the  \\ 
        & $r$-band reference image (deg) &  & $z$-band reference image (deg) \\
        r\_dec  & Declination measured on the $r$-band         & z\_dec  & Declination measured on the $z$-band \\
        & reference image (deg) & & reference image (deg) \\
        r\_mag  & Approximately calibrated $r$-band & z\_mag  & Approximately calibrated $z$-band \\
        & magnitude (AB) & & magnitude (AB) \\
        r\_merr & $r$ magnitude uncertainty                   & z\_merr & $z$ magnitude uncertainty \\
        \hline
        \multicolumn{4}{c}{\textbf{Figures}} \\
        \multicolumn{4}{c}{\textit{(Pre‑generated figures for quick check.)}} \\
        \hline\hline
        Type & Description & Type & Description \\
        \hline
        finding chart & The $1'\times1'$ finding chart centered on & CMD & The $(r-z)$ versus $z$ CMD of all sources \\
        & the target star & & located in the same $2.6'\times2.6'$ stamp field \\
        & & & as the target; target star highlighted \\
        \hline
    \end{tabular}
    \label{tab:data_products}
\end{table*}

\subsection{Web Portal Functionality}
For DR1, the DREAMS data portal is designed to enable exploration and retrieval of individual sources. The portal offers two primary functions:

\textbf{Source search.} Users can search for sources by coordinates (RA, Dec) or by DREAMS source identifier. The search returns a list of matching sources with basic information (coordinates, magnitudes) and links to detailed preview pages.

\textbf{Source preview and download.} For a selected source, the preview page displays:
\begin{itemize}
    \item A finding chart.
    \item A CMD with the source highlighted.
    \item The full $z$- and $r$-band light curves, with options to toggle between flux and magnitude. Bad-quality epochs (flagged by $\sigma_\mathrm{subt}$ or/and $\sigma_\mathrm{phot}$) are labeled.
\end{itemize}
From this page, users can download all associated data for the source, including the finding chart, CMD information, and light curves. Table~\ref{tab:data_products} lists the detailed description of all provided information for the single source.

\subsection{Current Release and Future Plans}
The current release (DR1) is optimized for targeted investigations: users can query and retrieve light curves for specific known sources or small samples. This enables, for example, follow-up studies of previously identified variables, rapid characterization of transient alerts from other surveys, and detailed analysis of individual objects of interest.

For large-scale blind searches, i.e., systematic variability or transient search across all sources in DREAMS field, one would require access to the full light curve database and a deeper, more complete input catalog without duplicates. 
This capability is planned for a future data release, anticipated after the 2026 observing season, following an update to the input catalog that incorporates deeper reference observations (Section~\ref{sec:dis}).

\section{Discussion}\label{sec:dis}

\begin{figure}
    \centering
    \includegraphics[width=1.0\linewidth]{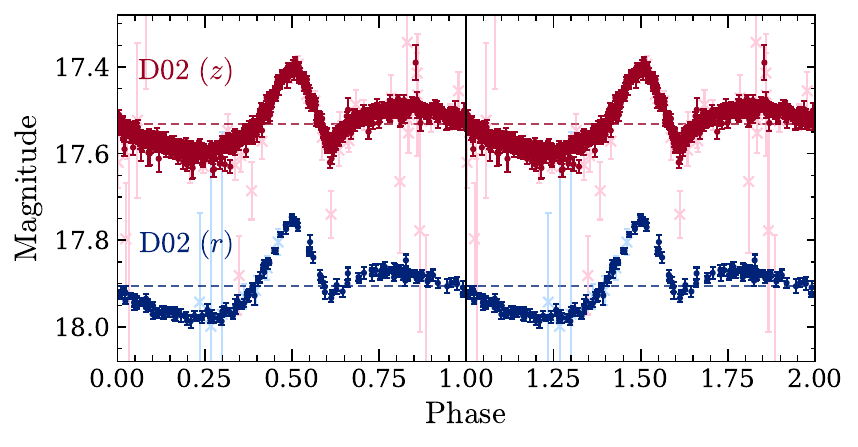}
    \caption{DREAMS $z$- and $r$-band light curves of the known BLAP OGLE-BLAP-019 \citep{Pietrukowicz2025_BLAP_OGLE}, which has a period of $\sim$48 min. Data points of poor quality are shown with reduced opacity.}
    \label{fig:lc_blap}
\end{figure}

\begin{figure}
    \centering
    \includegraphics[width=0.90\linewidth]{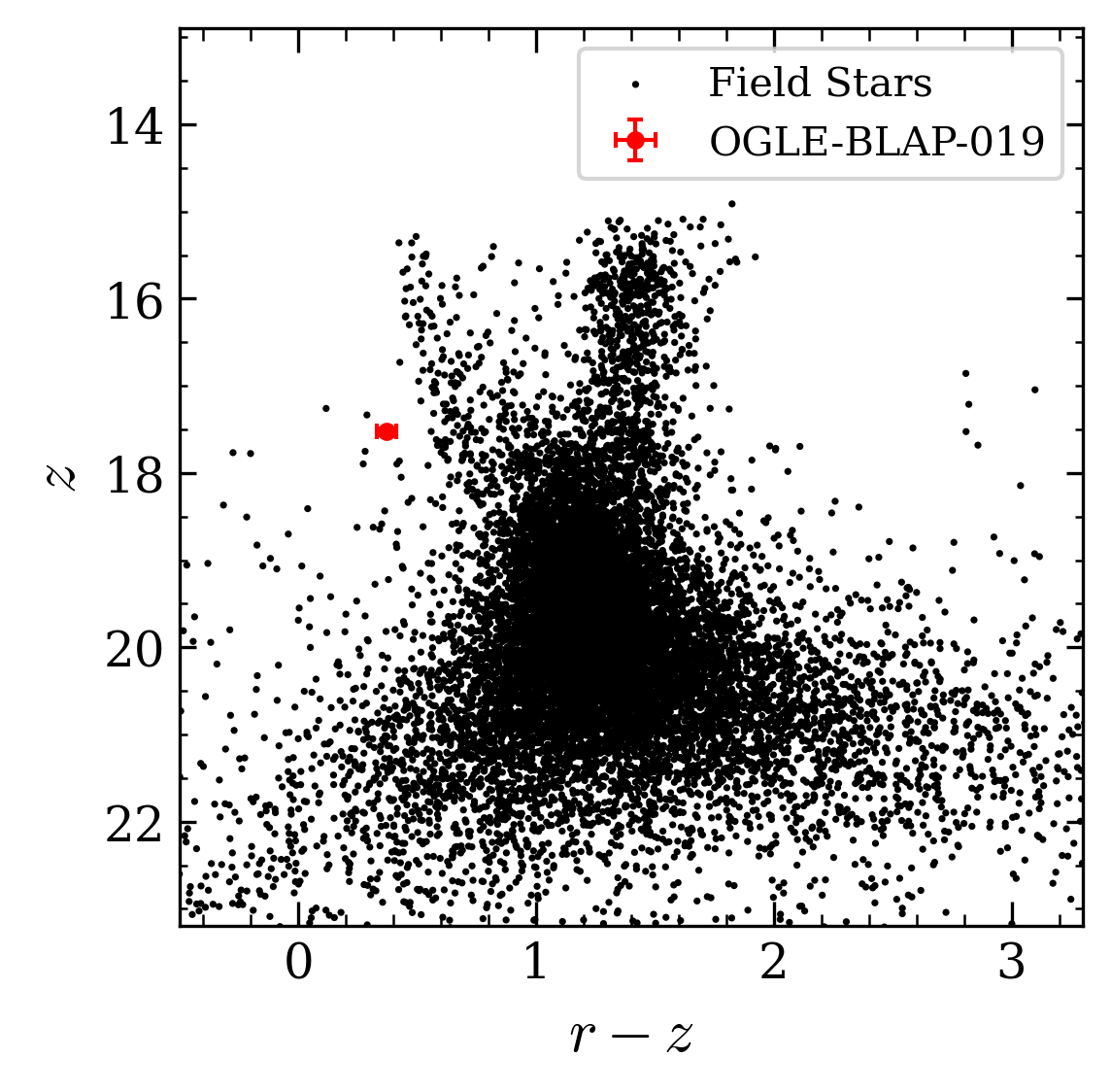}
    \caption{DREAMS $r-z$ and $z$ color-magnitude diagram of the known BLAP, OGLE-BLAP-019 \citep{Pietrukowicz2025_BLAP_OGLE}. Field stars are from the same 2.6$'\times$2.6$'$ stamp on D01 field, and the target is marked in red. The CMD indicates that OGLE-BLAP-019 is a very blue star.}
    \label{fig:cmd_blap}
\end{figure}

\subsection{Example Light Curves for Known Short Variables}

To demonstrate the capabilities of DREAMS, we present the $z$- and $r$-band light curves of two known short-period variables.

The first is a blue large-amplitude pulsator (BLAP), a recently identified class of hot, short-period pulsating stars first established from OGLE data. The original OGLE BLAPs have periods of about $20$--$40$\,min and optical amplitudes of $0.2$--$0.4$\,mag, but the now enlarged BLAP sample (now numbering close to 200) spans a broader range, roughly a few minutes to more than an hour, with amplitudes typically of order $0.1$--$0.4$\,mag and occasionally up to about $0.45$\,mag \citep{Pietrukowicz2017_BLAP_OGLE,Pietrukowicz2025_BLAP_OGLE,Borowicz2025}. 
BLAPs are generally interpreted as evolved, stripped, or otherwise binary-processed hot stars (see below), but their formation channels is still not yet completely understood. Possible channels include low-mass helium-core pre-white dwarfs with residual hydrogen-shell burning, core-helium-burning stripped stars, shell-helium-burning hot subdwarfs, and merger products 
\citep{Romero2018,Byrne2018,ByrneJeffery2020,WuLi2018,Byrne2021,Xiong2022,Lin2023_BLAP,KolaczekSzymanski2024}. Their variability is usually attributed to radial pulsation driven by the $\kappa$-mechanism \citep{Byrne2018,ByrneJeffery2020,Jeffery2025}.

Figure~\ref{fig:lc_blap} shows the DREAMS $z$- and $r$-band light curves of OGLE-BLAP-019 \citep{Pietrukowicz2025_BLAP_OGLE}, which has $I \sim 17.2$\,mag, a period of $\sim 48$\,min, and lies within the DREAMS D02 field (DREAMS DR1 ID: 45011997). With only the DR1 data, DREAMS already densely covers the full period in the $z$ band and nearly fully covers it in the $r$ band. In addition, the CMD provided in the data release (Figure~\ref{fig:cmd_blap}) clearly shows that the star is a very blue star. This case also demonstrates that the continuous overnight monitoring of DREAMS can enable the determination of periods of short-period variables using single- or few-night data, thereby allowing investigations of period changes on timescales of days to months (e.g., \citealt{Borowicz2025}).

\begin{figure}
    \centering
    \includegraphics[width=1.0\linewidth]{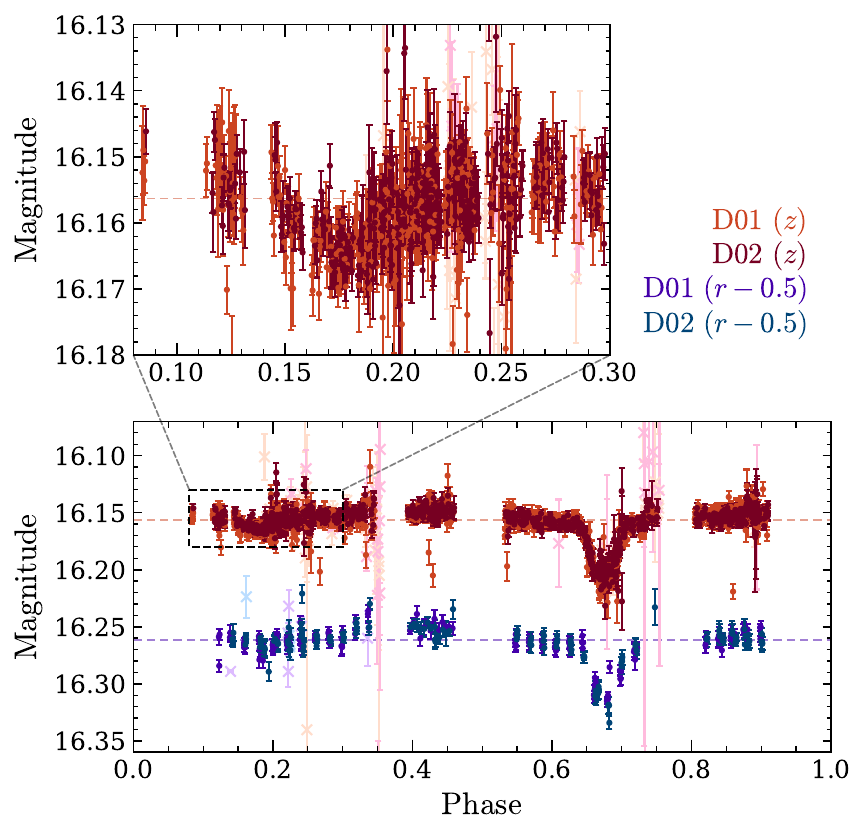}
    \caption{DREAMS $z$- and $r$-band light curves of the known transiting system OGLE-TR-18 \citep{OGLEIII-transit}. The DREAMS data newly reveal a secondary eclipse (shown in the upper panel) and phase-dependent brightness variations.}
    \label{fig:lc_tr}
\end{figure}

The second example is a transiting system identified by the OGLE-III survey, OGLE-TR-18, which has $I = 16.01$ mag, a period of 2.228 days, and a transit depth of $4.3\%$ \citep{OGLEIII-transit}. It lies within the DREAMS D01 field (DREAMS DR1 ID: 04230647), and its DREAMS light curves are shown in Figure~\ref{fig:lc_tr}. The DREAMS data recover the transit signal reported by OGLE. 
\HL{We have not performed a detailed quantitative comparison of the transit depth, as the primary goal is to demonstrate the data quality. However, by eye, the DREAMS $z$-band transit appears slightly deeper ($\sim 4.6\%$ or $\sim 0.05$\,mag) than the OGLE $I$-band value. Such color-dependent analysis could be of interest for other known transiting systems as well.}
In addition, DREAMS reveals a secondary eclipse with a depth of \HL{$\sim 0.7\%$ (or $\sim0.008$\,mag)} in the $z$ band, as well as phase-dependent brightness variations across the orbit. These features indicate that the system is most likely a detached eclipsing binary, with the companion being a low-mass stellar object, which is consistent with the results from the radial velocity follow-up \citep{OGLEIII-transit-followup}. This example demonstrates that DREAMS can verify previously reported transiting systems, identify new candidates, and, when combined with long-term monitoring from other surveys, enable studies of orbital evolution such as decay.

\subsection{A Pilot Search for Short Transient Signals}

To demonstrate the capability of DREAMS for detecting short-duration transient signals, we perform a pilot search over a small but representative region. This region is the overlapping area between the D01 N22 chip and the D02 S18 chip, covering about $0.04$\,deg$^2$. It contains 245,490 sources that are independently identified in both fields and were successfully cross-matched (Section~\ref{sec:pipeline:catalog}).

We search for short-duration brightening events among the 245,490 sources in the $z$-band light curves using a sliding box-fit algorithm. For each star, we combine all $z$-band data and scan a series of time windows with durations logarithmically spaced between 0.04 and 6 days, stepping by 1/10 of the window length. For each candidate window, we require at least five data points both inside and outside the window, and that at least five consecutive in-window points show significant residuals ($>3\sigma$) above a flat baseline. We then fit a two-level step function (constant flux inside the window and constant flux outside) and compute $\Delta\chi^2$ between this box model and a single flat model over the entire light curve. A window is flagged as a candidate if $\Delta\chi^2 > 300$ and the reduced $\chi^2$ of the out-of-window baseline is less than 3 to ensure the baseline itself is not highly variable. The window with the highest $\Delta\chi^2$ per detection is retained as the best candidate for each star.

This simple pilot search returns 97 candidate variables. After visual inspection, we identify two stellar flare events, one new microlensing event, and 31 likely low-amplitude short variable stars. The remaining candidates are false positives caused by systematics such as bleeding from bright stars or contamination from nearby variables. 

\begin{figure}
    \centering
    \includegraphics[width=1.0\linewidth]{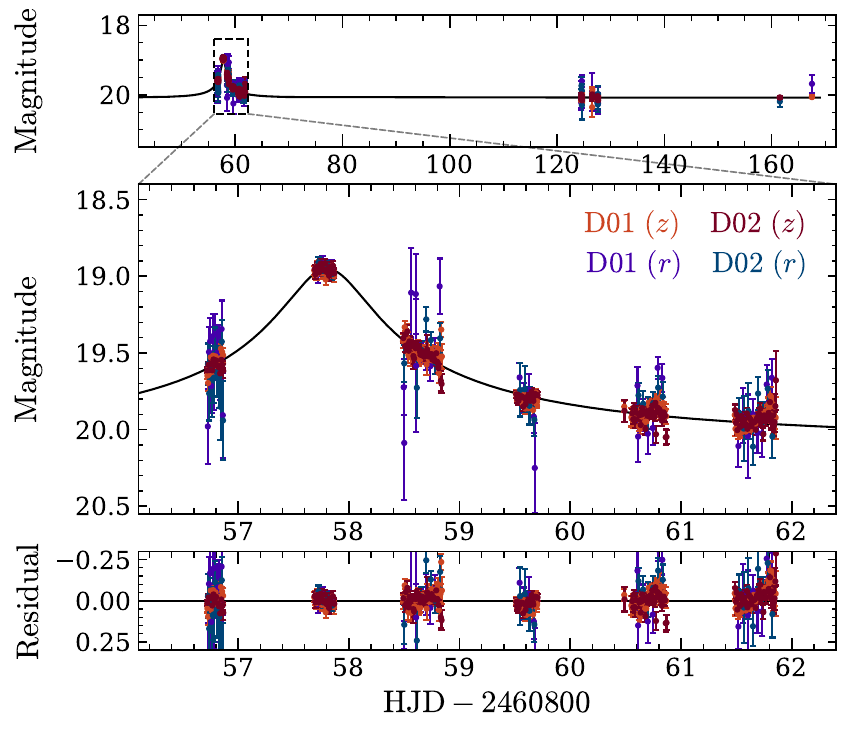}
    \caption{$z$- and $r$-band light curves of the newly discovered microlensing event, DREAMS-2025-BLG-0001, on the DREAMS DR1 source 03239519. For clarity, data in each observation block are binned. All magnitudes are aligned to the D01 $z$ band data following the microlensing model.}
    \label{fig:lc_ulens}
\end{figure}

Figure~\ref{fig:lc_ulens} displays the light curve of the newly discovered microlensing event located at the position of DREAMS DR1 star 03239519, with equatorial coordinates $(\alpha, \delta)_{\text{J2000}}$ = (17:53:57.45, $-$29:26:05.99) and Galactic coordinates $(\ell, b)$ = (0$^\circ$.5065, $-$1$^\circ$.8176). We designate this event as DREAMS-2025-BLG-0001, following the standard naming convention. A point-source point-lens (PSPL) model \citep{Paczynski1986} fit yields a microlensing timescale of $\tE = 9.0 \pm 1.1$ days and indicates a very faint source star with $z_{\rm S} = 22.83 \pm 0.15$ mag and $r_{\rm S} = 25.13 \pm 0.15$ mag. 
\HL{The baseline magnitude in Figure~\ref{fig:lc_ulens} is significantly brighter than the source because it includes flux from unresolved blended stars (e.g., the lens, a companion of the lens or source star, or nearby field stars), which is a common situation in microlensing.}

\begin{figure}
    \centering
    \includegraphics[width=1.0\linewidth]{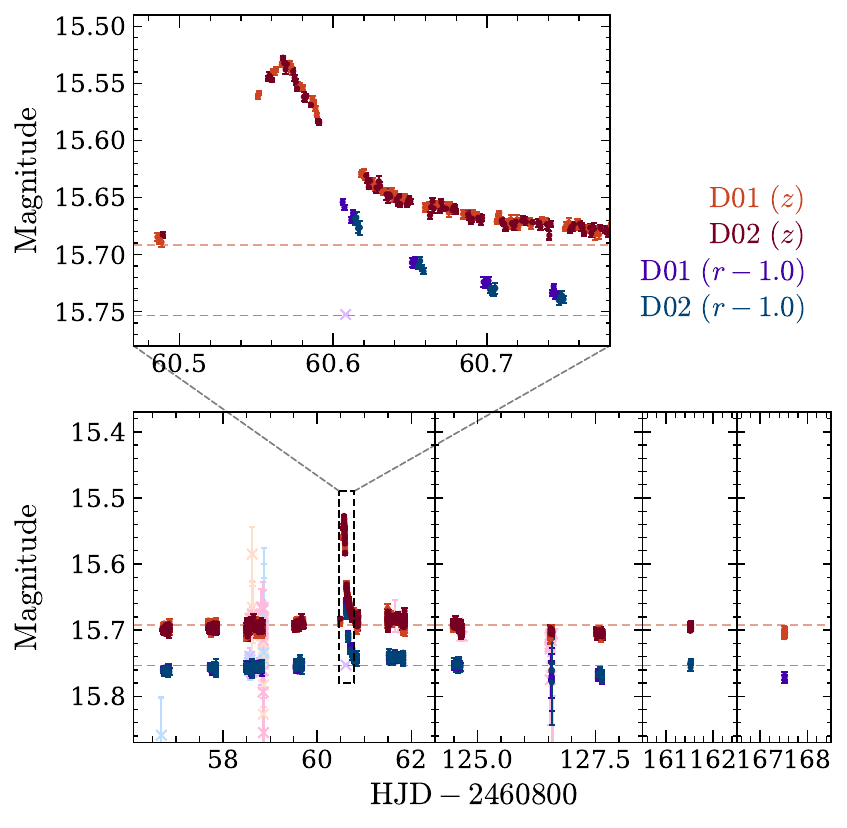}
    \caption{$z$- and $r$-band light curves of the stellar flare event (DREAMS DR1 ID 03032621) identified by a pilot search for short variables.}
    \label{fig:lc_flare}
\end{figure}

Figure~\ref{fig:lc_flare} shows the $z$- and $r$-band light curves of a short-duration flare on the DREAMS DR1 star 03032621 at equatorial coordinates $(\alpha, \delta)_{\text{J2000}}$ = (17:54:45.15, $-$29:32:25.07) and Galactic coordinates $(\ell, b)$ = (0$^\circ$.5030, $-$2$^\circ$.0201). Although there was a $\sim 1$\,hr gap during the rising side due to a technical issue, the minute-level cadence DREAMS data exhibit a typical stellar flare shape with a rapid rise and a slower decay, with an amplitude of $\sim0.15$ mag in $z$ band. 

We note that this region was not selected randomly. It was initially targeted because an obvious new microlensing event was identified during an early visual inspection of the data. Consequently, the event counts presented here should not be used to extrapolate global new microlensing event rates across the entire DREAMS field. However, the detection of two flares within this small area, which is likely to be much less affected by such selection bias, suggests that hundreds of flare events may be present across the full DR1. Because flares are an important source of contamination for FFP searches \citep{Mroz2017a}, the minute-level cadence of DREAMS data has the potential to enable detailed studies of flare morphology in the bulge field and better discrimination between flares and FFPs than current surveys. 

For the 31 identified short-period variables, we cross-match them with the OGLE-III and OGLE-IV Collections of Variable Stars \citep{OGLE4RRLyr, OGLE4RRLyr2019, OGLE4Cepheids, OGLE4Cepheids2020, OGLE4deltaScuti2020, OGLE4deltaScuti2021, OGLE4Heartbeat, OGLE4Heartbeat2, OGLE4ECL, OGLE3LPV, OGLE4Mira}. We find that seven of our sources correspond to previously known variables, leaving 24 newly identified short variables. The current pilot search is preliminary and covers only $\sim 0.4\%$ of DR1. With more sophisticated search algorithms, the full DR1 may contain thousands of previously unknown short-period variables, and the complete three-year DREAMS program will provide a powerful dataset for studies of short variability.

\subsection{Catalog Depth}\label{sec:dis:catalog}

\begin{figure}
    \centering
    \includegraphics[width=0.97\linewidth]{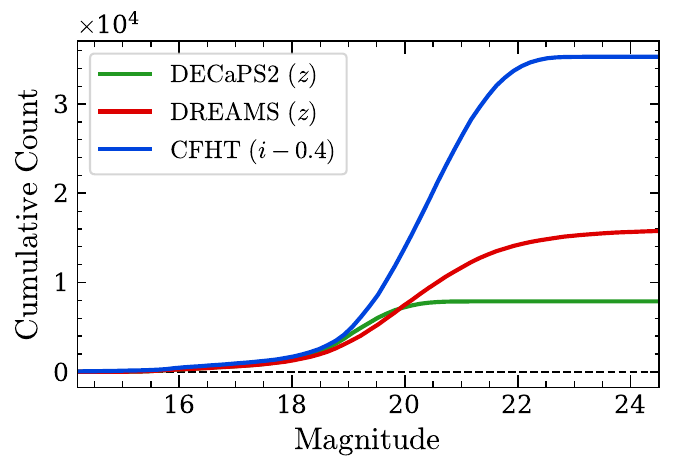}
    \caption{Cumulative number of catalog stars as a function of magnitude for DREAMS DR1 (red), DECaPS2 (green), and CFHT (blue), measured in a $2.53' \times 2.40'$ region centered at $(\alpha, \delta)_{\rm J2000}$ = (17:54:47.80, $-$29:32:50.27). The CFHT $i$-band magnitude are approximately converted to the $z$ band using a constant offset of 0.4 mag.}
    \label{fig:cat}
\end{figure}

The Dark Energy Camera Plane Survey 2 (DECaPS2; \citealt{DECaPS, Saydjari2023ApJS_DECaPS2}) covers the region $|b| \leq 10^\circ$ and $-124^\circ < \ell < 6^\circ$ in the $g$, $r$, $i$, $z$, and $Y$ bands. Each field typically has three visits per band, with a single-exposure time of 30\,s in the $z$ band. In contrast, the DREAMS DR1 reference image is constructed from a stack of 20 exposures, each with an exposure time of 42 or 60\,s, and is therefore expected to be deeper than the DECaPS2 catalog. 

Figure~\ref{fig:cat} shows the cumulative number of catalog stars as a function of magnitude for a $2.53' \times 2.40'$ region centered at $(\alpha, \delta)_{\rm J2000}$ = (17:54:47.80, $-$29:32:50.27), located on the D01 N22 chip. Compared to DECaPS2, the DREAMS catalog contains about twice as many stars, with most of the additional sources at $z > 20$ mag. To avoid duplicates in the DREAMS catalog, we only show stars detected by D01 reference images without any merging. 

Interestingly, the DECaPS2 catalog includes slightly more stars in the range $18.5 \lesssim z \lesssim 19.5$ mag. This difference likely arises for two reasons. First, the PSF FWHM of the DECaPS2 reference images is $\sim 1.0''$, compared to $\sim 0.85''$ for DREAMS. Second, \citet{Saydjari2023ApJS_DECaPS2} used the \texttt{crowdsource} package \citep{crowdsource} for star extraction, which tends to treat closely spaced stars as a single object when fainter stars are below a detection threshold. As a result, stars in the DECaPS2 catalog are more blended.

\HL{We also examined the catalog from the VISTA Variable in the Via Lactea Survey \citep[VVV,][]{VVV, Smith+2025_VVV2025} in the same region; it contains only $\sim 10^3$ sources with its $Z$-band photometry. This is consistent with the shorter exposure and narrower filter of VVV. Thus, it is not included in the comparison in Figure~\ref{fig:cat}. }

\HL{In the same small region, CFHT has archival observations, so we include a comparison here.} Figure~\ref{fig:cat} also shows the cumulative number of stars from a catalog constructed for the same field using $i$-band images obtained with CFHT between 2018 and 2022. The reference image is created by stacking 10 exposures, each with an exposure time of 10\,s, and has a PSF FWHM of $\sim 0.6''$. The CFHT catalog is constructed using the \textsc{DoPHOT} package. Owing to the better seeing, the CFHT catalog contains about 2.2 times more stars than the DREAMS catalog. However, the CFHT images do not provide full coverage of the DREAMS field, and the data, taken 3--7 years earlier, exhibit astrometric offsets relative to the DREAMS images. Therefore, the CFHT catalog is not incorporated into DR1. We plan to construct a deeper reference catalog for DREAMS using CFHT or Subaru in 2026 and apply it in DREAMS DR2.

\begin{acknowledgments}
H.Y. acknowledge support by the China Postdoctoral Science Foundation (No. 2024M762938). H.Y., W.Z., Q.Q., Y.T., Z.L., Y.S., H.M., J.Z., H.L., X.S., and S.M. acknowledge support by the National Natural Science Foundation of China (Grant No. 12133005, PI: S.M.). The authors acknowledge the High-performance Computing Center at Westlake University for providing computational and data storage resources that have contributed to the research results reported within this paper. The authors acknowledge the Office of Information Technology at Westlake University for helping with the data transfer and developing and deploying the data release site. 
This work is part of the ET space mission which is funded by the China's Space Origins Exploration Program. 
The work of K.B. is supported by NOIRLab, which is managed by the Association of Universities for Research in Astronomy (AURA) under a cooperative agreement with the U.S. National Science Foundation. 
This research was funded in part by National Science Centre, Poland, grant SONATA 2023/51/D/ST9/00187 awarded to P.M.
J.C.Y. acknowledges support from U.S. NASA Grant No. 80NSSC25K7146. 
R.A.S and K.K. acknowledge support from US National Science Foundation grant 2206828. 
T.C. was supported by NASA through the NASA Hubble Fellowship grant HST-HF2-51527.001-A awarded by the Space Telescope Science Institute, which is operated by the Association of Universities for Research in Astronomy, Inc., for NASA, under contract NAS5-26555.

This project used data obtained with the Dark Energy Camera (DECam), which was constructed by the Dark Energy Survey (DES) collaboration. Funding for the DES Projects has been provided by the U.S. Department of Energy, the U.S. National Science Foundation, the Ministry of Science and Education of Spain, the Science and Technology Facilities Council of the United Kingdom, the Higher Education Funding Council for England, the National Center for Supercomputing Applications at the University of Illinois at Urbana-Champaign, the Kavli Institute for Cosmological Physics at the University of Chicago, the Center for Cosmology and Astro-Particle Physics at The Ohio State University, the Mitchell Institute for Fundamental Physics and Astronomy at Texas A\&M University, Financiadora de Estudos e Projetos, Funda\c{c}\~ao Carlos Chagas Filho de Amparo \`a Pesquisa do Estado do Rio de Janeiro, Conselho Nacional de Desenvolvimento Cient\'{\i}fico e Tecnol\'ogico and the Minist\'erio da Ci\^encia, Tecnologia e Inova\c{c}\~ao, the Deutsche Forschungsgemeinschaft, and the collaborating institutions in the Dark Energy Survey.

The collaborating institutions are Argonne National Laboratory; the University of California at Santa Cruz; the University of Cambridge; Centro de Investigaciones Energ\'eticas, Medioambientalesy Tecnol\'ogicas (CIEMAT), Madrid; the University of Chicago; University College London; the DES-Brazil Consortium; the University of Edinburgh; the Eidgen\"ossische Technische Hochschule (ETH) Z\"urich; Fermi National Accelerator Laboratory; the University of Illinois at Urbana-Champaign; the Institut de Ci\`encies de l'Espai (IEEC/CSIC); the Institut de F\'{\i}sica d'Altes Energies (IFAE); Lawrence Berkeley National Laboratory; the Ludwig-Maximilians-Universit\"at M\"unchen and the associated Excellence Cluster Universe; the University of Michigan; NSF NOIRLab; the University of Nottingham; The Ohio State University; the OzDES Membership Consortium; the University of Pennsylvania; the University of Portsmouth; SLAC National Accelerator Laboratory; Stanford University; the University of Sussex; and Texas A\&M University.

Based on observations at NSF Cerro Tololo Inter-American Observatory, NSF NOIRLab (NOIRLab Prop.\ ID 2025A-806294, PI: Weicheng Zang; Prop.\ ID 2025B-560332, PI: Weicheng Zang \& Hongjing Yang), which is managed by the Association of Universities for Research in Astronomy (AURA) under a cooperative agreement with the U.S. National Science Foundation.
\end{acknowledgments}

\software{pySIS \citep{pysis,Yang2024_pysis5_RAMP1,Yang2025_RAMP2}, 
          DECam Community Pipeline \citep{DECampipeline},
          NumPy \citep{numpy:2020}, 
          Matplotlib \citep{Matplotlib}, 
          SciPy \citep{scipy:2020}, 
          H5py \citep{h5py}}

\bibliography{Yang.bib}{}
\bibliographystyle{aasjournalv7}

\end{CJK*}
\end{document}